# *mendelFix*: a Perl script for checking Mendelian errors in high density SNP data of trio designs

Yuri Tani Utsunomiya[1]*, Rodrigo Vitorio Alonso[2,3], Adriana Santana do Carmo[3], Francine Campagnari[3,4], José Antonio Vinsintin[2], José Fernando Garcia[5]

[1]Departamento de Medicina Veterinária Preventiva e Reprodução Animal. Faculdade de Ciências Agrárias e Veterinárias, Jaboticabal SP-Brazil. UNESP - Univ Estadual Paulista. [2]Departamento de Reprodução Animal. Faculdade de Medicina Veterinária e Zootecnia, São Paulo SP-Brazil. USP - Universidade de São Paulo. [3]Deoxi Biotecnologia Ltda., Araçatuba SP-Brazil. [4]Departamento de Genética. Instituto de Biociências, São Paulo SP-Brazil. USP - Universidade de São Paulo. [5]Departamento de Apoio, Produção e Saúde Animal. Faculdade de Medicina Veterinária de Araçatuba, Araçatuba SP-Brazil. UNESP - Univ Estadual Paulista

*Corresponding author: ytutsunomiya@gmail.com

**Abstract**

Here we present *mendelFix*, a Perl script for checking Mendelian errors in genome-wide SNP data of trio designs. The program takes 12-recoded PLINK PED and MAP files as input to calculate a series of summary statistics for Mendelian errors, sets missing offspring genotypes that present Mendelian inconsistencies, and implements a simplistic procedure to infer missing genotypes using parent information. The program can be easily incorporated in any pipeline for family-based SNP data analysis, and is distributed as free software under the GNU General Public License.

**Keywords**: SNP, Mendelian error, Perl

## Background

In recent years, the release of reference genome assemblies, together with initiatives for genome re-sequencing, has enabled the ascertainment of millions of single nucleotide polymorphisms (SNPs) with common variants across populations in several species of animals. In parallel, the development of high-throughput DNA microarray technologies allowed for the rise of large-scale genotyping, making it available huge amount of data for genome-wide SNP variation surveys. In spite of the high accuracy, reproducibility and automation achieved by these platforms, high-throughput genotyping is not immune to errors, and the size of data files makes the point-by-point assessment of data quality unfeasible.

In the particular case of investigations with family-based designs, pedigree data and genotypes of parents and offspring are important components of genomic datasets, and one essential quality parameter that can be controlled in this setting is the consistency between the genotypes of the offspring and the genotypes that are observed in the parents, according to Mendel's First Law (Law of Segregation) [1]. The detection of such genotyping errors is crucial before any downstream analysis, and error summary statistics can be used to identify unreliable samples or markers, being helpful to decide whether a particular SNP or individual should be removed from the dataset [2]. A number of widely known programs offers means to calculate summary statistics for Mendelian errors and uses them as inclusion criteria for both SNPs and individuals, such as *Illumina®GenomeStudio/BeadStudio* [3], *PLINK* [4] and *Golden HelixSNP & Variation Suite* (*SVS*) [5].

Beyond reporting summary statistics and exclude SNPs and individuals with high error rates, one may be willing to set these errors to missing, and use the Law of Segregation to infer unobserved genotypes in the offspring. Here, we present *mendelFix*, a simple Perl script that can cope with both tasks: i. calculate a series of statistics for Mendelian inconsistencies and ii. correct offspring genotypes based on parent information. The script is simple and can be easily incorporated in the pipelines of users that are familiar with *PLINK*.

**Description of the script**

The script assumes that the parentage information provided in the pedigree data and the genotypes of the parents are correct. Hence, in order to obtain reliable results, it is recommended that the user certifies that correct pedigree information and high quality parental genotype data are provided. The script is entirely based on the following rationale: as an individual is expected to share one allele identical by descent (IBD) with each one of its parents at any given locus, it is also expected that this individual shares one or two alleles identical by state (IBS) with each one of its parents at the same locus. This assumption, based on Mendel's First Law, is then used to derive a series of summary statistics.

Let IBS0, IBS1 and IBS2 denote cases where the comparison of the genotypes of two individuals for a given SNP results in zero, one or two alleles identical by state, respectively. The number of Mendelian errors between parent and offspring is the simple count of IBS0. The proportion of IBS0 across all comparable SNPs for a parent-offspring pair is referred as the Parent-Child error (PC-error). Note that the PC-error reflects the error rate when a single parent is observed, and can be divided into Father-Offspring error and Mother-Offspring error.

In the case where both parents are being considered simultaneously, the problem can be restated as follows. Let 1 and 2 denote the states of two alleles in a given SNP. If both parents are 12 (i.e., heterozygotes), then the offspring could assume any genotype class - 11, 12 or 22. In this case, incorrectly assigned genotypes are not detectable by means of checking Mendelian errors. However, if at least one of the parents is homozygous for one of the alleles, there are genotype classes that cannot be present in the offspring according to Mendelian inheritance, and an error can be easily detected. This will result in either the appearance of an allele that should not be present in the genotype of the offspring, namely Allele Drop In (ADI), or an allele that should be present, but is absent, in the genotype of the offspring, namely Allele Drop Out (ADO). **Table 1** summarizes all possibilities for ADI and ADO occurrences. This type of error can only be checked if genotypes are observed for all members of the trio. The ratio between ADI or ADO and either the total number of calls in the offspring or the number of comparable SNPs for a father-mother-offspring trio, are useful indicators to assess genotyping quality. The sum of the occurrences of ADI and ADO reflects the total number of Mendelian errors in the offspring, and the proportion of Mendelian errors across all comparable SNPs in a trio is referred as the Parent-Parent-Child error (PPC-error) or Father-Mother-Offspring error.

The main goal of *mendelFix* is to calculate all the above statistics for each offspring, set offspring genotypes that present Mendelian inconsistencies as missing, and then infer missing genotypes using parent information. This inference is based on a simple procedure: if both parents are homozygous for the same allele, then the same genotype is assigned to the offspring; else, if the parents are homozygous for different alleles, it is assigned a heterozygous genotype to the offspring. When at least one of the parents is heterozygous for a given SNP, estimating the genotype of the offspring may require the use of phased data and methods of inference that account for haplotype structure, which are not currently implemented in the program. Hence, *mendelFix* offers a simplistic genotype inference for cases where both parents are homozygous, and can be used as an auxiliary tool preceding imputation analyses.

**Summary statistics calculated**

For each individual with parental genotypes available, *mendelFix* calculates the following statistics:

*Total number of Calls (NCALL)*. The absolute number of observed genotypes in the offspring. Two versions of this statistic are provided: *NCALL1*, representing the number of Calls before fixing genotypes, and *NCALL2*, the number of Calls after fixing genotypes.

*Call rate (CR)*. Ratio between *NCALL* and the total

number of SNPs. This statistic is also provided in two versions: *CR1* and *CR2*, for *NCALL1/NSNP* and *NCALL2/NSNP*, respectively.

*Number of comparable calls between father and offspring (FOCALL)*. Number of SNPs where genotypes are observed in both father and offspring.

*Father-Offspring IBS0 (FOIBS0)*. Number of cases where the comparison of the genotype of the father and the genotype of the offspring result in zero alleles identical by state.

*Father-Offspring IBS1 (FOIBS1)*. Number of cases where the comparison of the genotype of the father and the genotype of the offspring result in one allele identical by state.

*Father-Offspring IBS2 (FOIBS2)*. Number of cases where the comparison of the genotype of the father and the genotype of the offspring result in two alleles identical by state.

*Father-Offspring error (FOERROR)*. Ratio between *FOIBS0* and *FOCALL*.

*Number of comparable calls between mother and offspring (MOCALL)*. Number of SNPs where genotypes are observed in both mother and offspring.

*Mother-Offspring IBS0 (MOIBS0)*. Number of cases where the comparison of the genotype of the mother and the genotype of the offspring result in zero alleles identical by state.

*Mother-Offspring IBS1 (MOIBS1)*. Number of cases where the comparison of the genotype of the mother and the genotype of the offspring result in one allele identical by state.

*Mother-Offspring IBS2 (MOIBS2)*. Number of cases where the comparison of the genotype of the mother and the genotype of the offspring result in two alleles identical by state.

*Mother-Offspring error (MOERROR)*. Ratio between *MOIBS0* and *MOCALL*.

*Total number of comparable SNPs among all members of the trio (TRIOCALL)*. The absolute number of SNPs for which determined genotypes are observed for the mother, the father and the offspring.

*Number of occurrences of Allele Drop In (ADI)*. Number of cases where the offspring presents an allele that could not have been inherited from its parents.

*ADI/TRIOCALL ratio (PADI)*. The proportion of the comparable genotypes that resulted in ADIs.

*Number of occurrences of Allele Drop Out (ADO)*. Number of cases where the offspring do not present an allele that should have been inherited from its parents.

*ADO/TRIOCALL ratio (PADO)*. The proportion of the comparable genotypes that resulted in ADOs.

*NERROR*. The absolute number of Mendelian errors among the SNPs for which determined genotypes are observed for the mother, the father and the offspring. This statistic is represented by the sum between ADI and ADO.

*Mother-Father-Offspring error or Parent-Parent-Child error (PPCERROR)*. Ratio between NERROR and TRIOCALL.

*Total number of fixed genotypes (NFIX)*. The number of missing genotypes that could be inferred by using parent information.

**Table 1**. Detectable Mendelian errors in bi-allelic SNP in trio designs.

| Allelic Drop | Father's genotype | Mother's genotype | Offspring's genotype |
|---|---|---|---|
| In (ADI) | 11 | 11 | 12/22 |
|  | 22 | 22 | 12/11 |
| Out (ADO) | 11 | 12/22 | 22 |
|  | 12/22 | 11 | 22 |
|  | 22 | 12/11 | 11 |
|  | 12/11 | 22 | 11 |

**Script workflow**

The script takes 12-recoded *PLINK* PED and MAP files [3] as input, and reads in the map and pedigree information provided. At this stage, *mendelFix* checks if the user has specified a set of chromosomes that should not be checked for Mendelian errors (optional). Individuals with missing parents, either by explicit declaration with missing values or by absence of the genotypes of a declared parent, are not analyzed. These individuals are listed in a file with a *.pmiss* suffix. Next, the main algorithm is executed in two steps. First, all summary statistics are calculated and offspring genotypes with Mendelian errors are set as missing. Second, parent information is used to infer missing genotypes in the offspring, according to the Law of Segregation. The summary statistics are then outputted to a file with the *.stats* suffix, and fixed genotype data are written to a new pair of PED and MAP files. The script has a series of internal controls for checking the data structure of the input files and arguments provided, causing the task to be aborted if unexpected formatting is found.

**Implementation and usage**

The application was implemented as a cross platform standalone Perl script, and was tested and validated using Perl v5.10.1 and v5.14.2 built for x86_64-linux-gnu-thread-multi [6], under Ubuntu Server 10.04 LTS (Lucid Lynx) and Ubuntu Desktop 12.04 LTS (Precise Pangolin) [7]. Basically, the code was written in four simple subroutines: *getMap()*, responsible for parsing the MAP file; *getIDs()*, which parses the pedigree information within the PED file; *getGenotypes()*, the genotype parser; and *mendelFix()*, the subroutine that contains the main algorithm. The program is free software under the terms of the GNU General Public License as published by the Free Software Foundation, either version 3 or any later version [8], and comes without any warranty. The complete script source code, tutorial and example files were made available at SourceForge (see *Availability and requirements*).

The usage of *mendelFix* is straightforward: *perl mendelfix.pl data.ped data.map outfile*. Optionally, chromosomes that should not be checked can be specified by simply adding their identifiers after the command line. For instance: *perl mendelfix.pl data.ped data.map X Y* will cause the script to not perform Mendelian checks on chromosomes X and Y. For further details and examples, see the documentation.

**Conclusions**

For those who wish to correct SNP data of trio designs for Mendelian errors instead of only calculating summary statistics, *mendelFix* is a useful, simple-to-use tool. The use of the widely known PED/MAP format as input/output facilitates the incorporation of *mendelFix* into pipelines for SNP analysis.

**Availability and requirements**

- Project name: mendelFix
- SourceForge link: http://sourceforge.net/projects/mendelfix
- Operating systems: cross platform
- Programming Language: Perl
- Requirements: Perl 5 or higher
- License: GNU General Public License (GPL) v3

**Author contributions**

JFG coordinated the research. RVA conceived the application. YTU implemented the code and prepared the manuscript. YTU and ASC tested the program. FC and JAV contributed to the discussion of the application. All authors revised and approved the final manuscript. The authors declare no competing interest.

**Acknowledgements**

This research was supported by: Conselho Nacional de Desenvolvimento Científico e Tecnológico (CNPq) – process 350291/2012-8, 551809/2011-6, 560922/2010-8 and 483590/2010-0; and Fundação de Amparo à Pesquisa do Estado de São Paulo (FAPESP) - process 2011/16643-2, 2010/52030-2, 2009/51784-6 and 2006/56683-5.